\documentclass{aa}
\usepackage{graphicx}
\usepackage{subfigure}

\newcommand{\be}{\begin{equation}}
\newcommand{\ee}{\end{equation}}
\newcommand{\bea}{\begin{eqnarray}}
\newcommand{\eea}{\end{eqnarray}}
\newcommand{\mb}[1]{\mbox{\boldmath$#1$}}
\newcommand{\mr}[1]{\mathrm{#1}}            
\newcommand{\lb}{\left}                     
\newcommand{\rb}{\right}

\begin{document}
\title{Radio emission and particle acceleration 
in plerionic supernova remnants}
\author{C. Nodes, G.T. Birk, M. Gritschneder, H. Lesch}
\institute{Institute for Astronomy and Astrophysics, University of Munich,
Germany}
              
\date{}
\mail{nodes@usm.uni-muenchen.de}

\abstract{Plerionic supernova remnants exhibit radio emission with remarkably
flat spectral indices ranging from $\alpha=0.0$ to $\alpha=-0.3$. The origin
of very hard particle energy distributions still awaits an explanation,
since shock waves generate particle distributions with synchrotron spectra
characterized by $\alpha\le-0.5$. Acceleration of high energy leptons in
magnetohydrodynamic turbulence instead may be responsible for the observed
hard spectra. This process is studied by means of relativistic test particle
calculations using electromagnetic fields produced by three-dimensional
simulations of resistive magnetohydrodynamical turbulence.  The particles
receive power-law energy spectra $N(\gamma)\propto \gamma^{-s}$ with $s$
ranging from $1.2$ to $1.6$, i.e. particle spectra that are required to
explain the radio emission of plerions.
 } 
\authorrunning{Nodes et al.}

\titlerunning{Particle Acceleration In Plerionic Supernova Remnants}
 \maketitle

\section{Introduction}
\label{intro}

Filled-center supernova remnants (SNR), or plerions exhibit flat radio
spectra, with power-law indices $0\leq \alpha\leq 0.3$ (for $S_\nu\propto
\nu^{-\alpha}$) (Weiler \& Panagia \cite{weiler1}; Weiler \& Shaver
\cite{weiler2}). Such spectra require energy distributions of the electrons
$N(\gamma)\propto \gamma^{-s}$ with $s=1+2\alpha$ be in the range $1\leq
s\leq 1.6$.  The flatness of their radio spectra distinguishes plerions
from a typical shell-type SNR with a mean $\alpha \sim 0.5$ implying
$N(\gamma)\propto \gamma^{-2}$ (e.g. Green 1991). The origin of relativistic
electrons responsible for these unusual radio spectra of plerions is not
yet understood (e.g. Green \cite{green2}; Woltjer et al. \cite{woltjer};
Arons \cite{arons}).

Although the many details of the nature and structure of plerions are
still unclear (see Arons \cite{arons}; Salvati et al. \cite{salvati};
Bandiera \cite{bandiera1} for reviews) there is a consensus on the following
scenario: Plerions are expanding bubbles, filled with magnetic fields and
relativistic leptons. Both components are continuously supplied to the
nebula by some central source, i.e. by a rotating neutron star in form of
a strongly magnetized wind whose energy is at least partially dissipated in
termination shock waves (Kennel \& Coroniti \cite{kennela,kennelb}; Galant \&
Arons \cite{galant}).

Arons (\cite{arons}) discussed the difficulties of particle acceleration in
plerions in great detail. With respect to prototypic Crab nebula he pointed
out that the emission at high frequencies (from the optical towards X- and
$\gamma$-rays) diagnoses the coupling physics {\it today} since the synchrotron
loss times of the high energy particles are significantly smaller than the
lifetime of the nebula.  The radio emission instead measures the integral
of the pulsars input over the entire history of the nebula - most of the
stored relativistic energy is in magnetic fields and radio emitting particles
(about $10^{50} ergs$).  Arons noted that averaged over the whole nebula,
the radio emitting spectrum has the form $N(\gamma)\propto \gamma^{-1.5}$,
$10^{2.5}\leq\gamma\leq 10^4$ based on the detailed spectral index maps by
Bietenholz \& Kronberg (\cite{bietenholz}) which show the particle distribution
to be remarkably homogeneous. The origin for this energy distribution is
not clear since the wind termination shock wave models do not yield power
laws flatter than $N(\gamma)\propto \gamma^{-2}$ at energies small compared
to $\gamma\sim 10^6$.  Arons speculated that some additional acceleration
physics by magnetohydrodynamic (MHD) waves may have to be included.

Bandiera et al. (\cite{bandiera2}) mapped the Crab Nebula at 230 GHz and
compared it 1.4 GHz map. The spectral index in the inner region at 230 GHz
is flatter (by $\sim 0.05$) than in the rest of the nebula. Furthermore they
found a steepening of the spectrum at the places of radio emitting filaments,
concluding that the magnetic field strengths in the filaments is higher than
in the surrounding nebula.  Some evidence for in-situ acceleration in the
Crab nebula has been collected by the radio observations of Bietenholtz,
Frail \& Hester (2001).

It is the aim of our contribution to show that resistive magnetohydrodynamical
fluctuations excited to a highly turbulent level accelerate test particles
to energy distributions with power law indices $1\leq s \leq 1.6$ being in
good agreement with the radio observations of plerions. We also show that
an increase in particle density as can expected for the filaments lead to
a steepening of the particle spectrum. Given the fact, that central sources
energize the nebula, thereby exciting strong MHD fluctuations we think that
our simulations may shed some light on the probably necessary additional
acceleration physics as speculated by Arons (\cite{arons}).

In the next section we describe the MHD simulations.  Section \ref{particle}
contains the results of relativistic particle simulations including radiative
synchrotron losses and finally we discuss our findings in Section 4

\section{Reconnective MHD turbulence}
\label{mhd}

Pulsars permanently power their environment by radiation and, in particular,
pulsar plasma winds. In fact, in the case of the Crab nebula the permanent
energy input by plasma flow is about $10^{38}$ erg/s. This plasma flow
encounters the termination shock at radius $R_\mr{s} \approx 10^{17} \mr{cm}$
where it gets decelerated and leptons are accelerated to ultra-relativistic
energies. Most of their synchrotron radiation is emitted in the optical to
$\gamma$-ray band. The presence of this termination shock is accompanied by
strong excitation of Alfv\'en waves which are expected to excite strong MHD
turbulence in the pulsar's nebula (Kronberg et al. \cite{kronberg}). In
this picture the size of the turbulence cells in the nebula is limited
by $R_\mr{s}$. The projected magnetic field within the Crab nebula was
found to have a coherence length of about $10^{16}\, cm$ (Bietenholz \&
Kronberg 1992). Kronberg et al (1993) suggested that the fact that this
length scale is close to the inner shock radius $R_\mr{s}$, may be more
than coincidental.  Laboratory studies of turbulent flows have revealed
the existence of persistence structures, which are advected in the flow and
exhibit sizes comparable to the source size (Hussein \cite{hussein1,hussein2}).
In other words, the source of turbulence within a supernova remnant is given
by a shock wave, like in the Crab nebula, the spatial scale of the shock is
the maximum length scale for turbulence within the nebula. To be precise,
in the Crab nebula the characteristic size of the turbulence should be
smaller than $10^{17}\mr{cm}$.

The associated MHD turbulence can be numerically generated by means of the
so-called Orszag-Tang turbulence (Orszag \& Tang \cite{orszag}). The
Orszag-Tang initial condition is a generic way to excite turbulence in a
magnetized plasma and is given by the non-linear interaction of Alfv\'en
waves which are MHD eigenmodes of a magnetized fluid e.g.,
\begin{eqnarray}
B_x &=& -B_0
   \sin\lb[2\pi (y-y_{{\rm min}})/(y_{{\rm max}}-y_{{\rm min}})\rb]\nonumber \\
 &&\sin\lb[2\pi (z-z_{{\rm min}})/(z_{{\rm max}}-z_{{\rm min}})\rb] \\
B_y &=& B_0
   \sin\lb[4\pi (x-x_{{\rm min}})/(x_{{\rm max}}-x_{{\rm min}})\rb]\nonumber \\
 &&\sin\lb[2\pi (z-z_{{\rm min}})/(z_{{\rm max}}-z_{{\rm min}})\rb] \\
B_z &=& B_0
   \sin\lb[4\pi (x-x_{{\rm min}})/(x_{{\rm max}}-x_{{\rm min}})\rb]\nonumber \\
 &&\sin\lb[2\pi (y-y_{{\rm min}})/(y_{{\rm max}}-y_{{\rm min}})\rb] \\
v_x &=&-v_0
   \sin\lb[2\pi (y-y_{{\rm min}})/(y_{{\rm max}}-y_{{\rm min}})\rb]\nonumber \\
 &&\sin\lb[2\pi (z-z_{{\rm min}})/(z_{{\rm max}}-z_{{\rm min}})\rb] \\
v_y &=& v_0
   \sin\lb[2\pi (x-x_{{\rm min}})/(x_{{\rm max}}-x_{{\rm min}})\rb]\nonumber \\
 &&\sin\lb[2\pi (z-z_{{\rm min}})/(z_{{\rm max}}-z_{{\rm min}})\rb] \\
v_z &=& 0 
\end{eqnarray}

The amplitudes are chosen as $B_0 = 5.5 \times 10^{-3}\;\mr{G}$ and
$v_0 = 0.05$, $0.016$, $0.005$, $0.0016$, $0.0005$, $0.00016 \;\mr{c}$ for
the six different test particle runs discussed in section \ref{particle}.

The nonlinear interaction of the Alfv\'enic perturbations result in almost
homogeneous turbulence.  We use periodic boundary conditions in all directions
and a numerical box given by $x_{\rm min}, y_{\rm min}, z_{\rm min}=
-0.5 \times 10^{16} \;\mr{cm}$ and $x_{\rm max}, y_{\rm max}, z_{\rm max}
= 0.5 \times 10^{16} \;\mr{cm}$ with a resolution of $101^3$ grid points.

We model this kind of turbulence by means of a well approved resistive
compressible 3D MHD code (Otto \cite{otto}). It integrates the balance
equations that govern the macroscopic low-frequency dynamics which read
\be
 {\partial  \rho\over \partial t} +\nabla\cdot\lb(\rho{\bf v}\rb)=0
\label{mhd1}\ee

\be
 {\partial  \rho {\bf v}\over \partial t}
 + \nabla\cdot\lb(\rho{\bf v}{\bf v}\rb) = - \nabla p
 + \frac{1}{4\pi}(\nabla \times {\bf B}) \times {\bf B}
\label{mhd2}\ee

\be
 {{\partial p} \over {\partial t}} =
 - {\bf{v}} \cdot \nabla p
 - \gamma p \nabla \cdot {\bf{v}}
 + \lb(\gamma - 1\rb) \eta \lb(\nabla \times\bf{B}\rb)^2
\label{mhd3}\ee

\be
 \frac{1}{c}{\partial {\bf B} \over \partial t} =
 \nabla \times \frac{1}{c}\lb({\bf v}\times {\bf B}\rb)
 - \frac{c}{4\pi}\nabla\times\lb(\eta\nabla\times{\bf B}\rb)
\label{mhd4}\ee
where $\rho$, ${\bf v}$, $p$, and ${\bf B}$ denote the mass density,
bulk velocity, thermal pressure, and the magnetic field.  By $\eta$ the
resistivity is denoted.

Reconnection is allowed due to a microturbulent resistivity which is modeled
in a current dependent way. The motivation for this kind of resistivity
is the following: In the highly turbulent almost ideal plasma of the
nebula collisional resistivity is negligible. Rather a wide variety of
micro-instabilities driven by electric currents is responsible for localized
dissipation. For strong turbulence the amplitude of the resistivity is given
by $\eta \approx m_e \omega_e / n e^2$, where $\omega_e=\sqrt{4\pi n e^2/m_e}$
is the electron plasma frequency (Vasyliunas \cite{vasyliunas}). This kind
of resistivity is assumed for our simulations.

\begin{figure}
  \resizebox{\hsize}{!}{
  \includegraphics{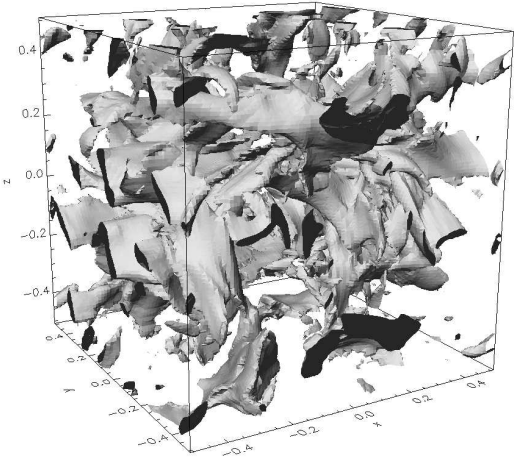}
  }
  \caption{3D Isosurface plot of the magnetic energy density. The shaded volume
  surface corresponds to a magnetic energy density of $u_\mr{magn} \approx 2.5
  \times 10^{-11} \mr{erg}/\mr{cm}^3$. The spatial extension is given in units
  of $10^{16}$ cm.}
  \label{b2_3d}
\end{figure}
\begin{figure}
  \resizebox{\hsize}{!}{
    \includegraphics{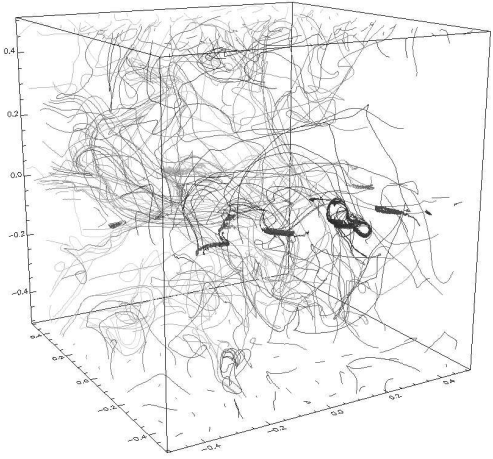}
  }
  \caption{Magnetic field lines.}
  \label{bflines}
\end{figure}
\begin{figure}
  \resizebox{\hsize}{!}{
    \includegraphics{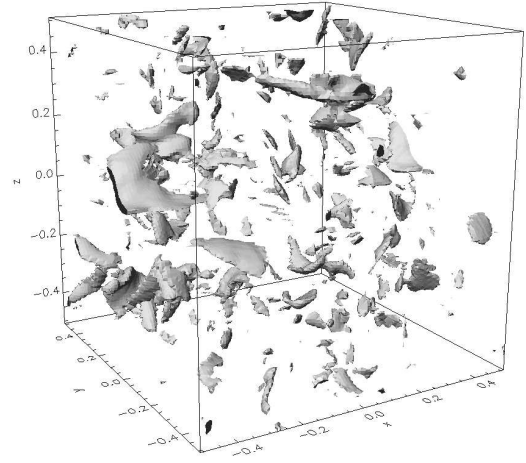}
  }
  \caption{Reconnection regions, i.e. regions with appreciable value of
  $E_{\|}$. The shaded volume surface corresponds to a value of $E_{\|}
  \approx 1.3 \times 10^{-6}, 4 \times 10^{-7}, 1.3 \times 10^{-7},
  4 \times 10^{-8}, 1.3 \times 10^{-8}, 4 \times 10^{-9} \; \mr{statvolt}
  /\mr{cm}$ for the six different test particle runs.}
  \label{epar}
\end{figure}

As a result of the resistive MHD calculations we get a turbulent field
configuration which we use for the test particle simulations. Fig. \ref{b2_3d}
and \ref{bflines} show the magnetic energy density and the magnetic field
lines.  These figures are to illustrate the turbulent structure
of the magnetic field. Fig. \ref{epar} shows the reconnection regions,
i.e. places where the parallel component of the electric field is appreciable
high. These are the acceleration sites where we expect the particles to gain
energy. They are also distributed in a very stochastic way clearly indicating
the turbulent nature of our fields.

\section{Relativistic particle simulations}
\label{particle}

The MHD calculations described in the previous section are used as the
electromagnetic environment for the studies of electron acceleration by means
of test particle simulations. We want to know how electric particles behave in
the complex three-dimensional turbulent electro-magnetic field configuration,
in particular how their energy distribution develops. 

The electric field is derived from the MHD quantities $\mb{B}$, $\mb{v}$ and
$\eta$ by means of the normalized Ohm's law
\be
  \label{ohm}
  \mb{E} = - \frac{1}{c}\mb{v} \times \mb{B} + \frac{c}{4\pi}
   \eta \mb{\nabla} \times \mb{B},
\ee
and gives for the mean value of of the electric field $\overline{E}\approx
0.02 \; \overline{B}$. Equation (\ref{ohm}) states that a parallel component
of the electric field $E_\|$ only occurs for $\eta \ne 0$. Compared to
the mean perpendicular component $\overline{E}_{\perp}$ the mean parallel
component in our simulations is found to be $\overline{E}_\| \approx 0.04 \;
\overline{E}_{\perp}$.

The results of the MHD calculation are scaled to model different physical
conditions, i.e. different mass densities inside the nebula, and are then used
for the test particle simulations. Here we present the results of particle
simulations for six different mean mass densities $\overline{\rho} = 10^{n}
m_p/cm^3$, $n=-2,-1,0,1,2,3$ (corresponding to the initial $v_0$). The
mean magnetic field strength for all runs was chosen to be $\overline{B} =
3 \times 10^{-4} \; G$.

Since the data for the electric and magnetic field is only available
on a discrete three-dimensional grid the test particle code uses linear
interpolation to determine the field values for any location. We note that
the term "test particle" means that the electromagnetic fields produced
by the particles are not changing the global fields, though the effect of
synchrotron radiation on the motion of the particles is taken into account,
i.e. the relativistic equations of motion have the form
\be
\frac{d\mb{p}}{dt} = q\left(\mb{E} 
+ \frac{1}{\gamma m c}\, \mb{p} \times \mb{B}\right) 
              + \mb{F}_{Rad}\;, \quad
\frac{d\mb{r}}{dt} = \frac{\mb{p}}{\gamma m},
\ee
where $m$ and $q$ denote the mass and charge of the particles, which are
electrons in our case and
\be
\gamma = \sqrt{1+\left(\frac{p}{mc}\right)^2}
\ee
is the Lorentz factor.

The calculations are relativistic, including the energy losses via synchrotron
radiation and inverse Compton scattering by (Landau \& Lifshitz \cite{landau})
\begin{eqnarray}
  \label{frad}
  \mb{F}_{Rad} 
  & \approx & \frac{2}{3} \frac{q^4}{m^2 c^4} 
  \lb\{ \mb{E} \times \mb{B} + 
  \frac{1}{c} \mb{B} \times \lb(\mb{B} \times \mb{v}\rb) + 
  \frac{1}{c} \mb{E} \lb(\mb{E\cdot v}\rb) \rb\}  \nonumber  \\
  & - &  \frac{2}{3} \frac{q^4}{m^2 c^5} \gamma^2 \mb{v} 
  \lb\{ \lb(\mb{E} + 
  \frac{1}{c} \mb{v} \times \mb{B}\rb)^2 - 
  \frac{1}{c^2} \lb(\mb{E\cdot v}\rb)^2 \rb\} 
\end{eqnarray}
The equations are numerically integrated by a Runge-Kutta algorithm of fourth
order with an adaptive stepsize control. As a result we get the momentum and
location at certain times of each particle in the given ensemble. The codes
has been used before to study high-energy particle acceleration (Nodes et
al. \cite{nodes}; Schopper et al. \cite{schopper}).

The direction of the initial momentum vector is equally distributed in a cone
of opening angle $80^{\circ}$ at the bottom of the box. This initial condition
corresponds to the idea that the electrons enter the the turbulent region
moving in a well defined direction. The notion "final state" relates to a set
of particles in which each individual particle has left the computational box.

The simulations start off with electrons injected into the
computational box at $z=z_{\rm min}$. We use a powerlaw $E_{kin}^{-2}$
ranging from $10^2$ to $10^4 \;m_ec^2$ as the initial energy spectrum for
all calculations. The choice of this injection spectrum was motivated by
our synchrotron model for the infrared to X-ray emission of the Crab pulsar
(Crusius-W\"atzel, Kunzl \& Lesch \cite{crusius}). There we could show
that the emission of the neutron star could be explained with a single
energy distribution $N(E)\propto E_{kin}^{-2}$. We could also show that
such an energy distribution is naturalley produced in an efficient pair
cascade. Given that plerions in general are powered by a pulsar like in
the case of the Crab nebula, an energy distribution $\propto E_{kin}^{-2}$
seems a quite reasonable choice as a probable injection spectrum in turbulent
plasmas of a plerionic supernova remnant.

Another series of simulations we performed showed clearly that the
resulting energy spectrum is independent of the index of the initial energy
distribution. Also the initial energy range does not play a crucial role for
the final distribution. After the initial injection of the electrons they get
accelerated very rapidly and their energy distribution reaches a stationary
state. This happens within only a few days and then the distribution remains
for several years until the final state is reached. For different initial
distributions only this first acceleration phase gets shorter or longer but
the final distribution is given by the energy offered by the MHD fields. Thus,
even for a non-relativistic initial energy distribution we expect the final
spectral index to be the same as the one found in the relativistic runs.

\begin{figure*}
  \centering
  \resizebox{\hsize}{!}{
    \subfigure[$\overline{\rho} = 0.01 \; m_p/cm^3$]
     {\includegraphics[width=8.5cm,keepaspectratio]{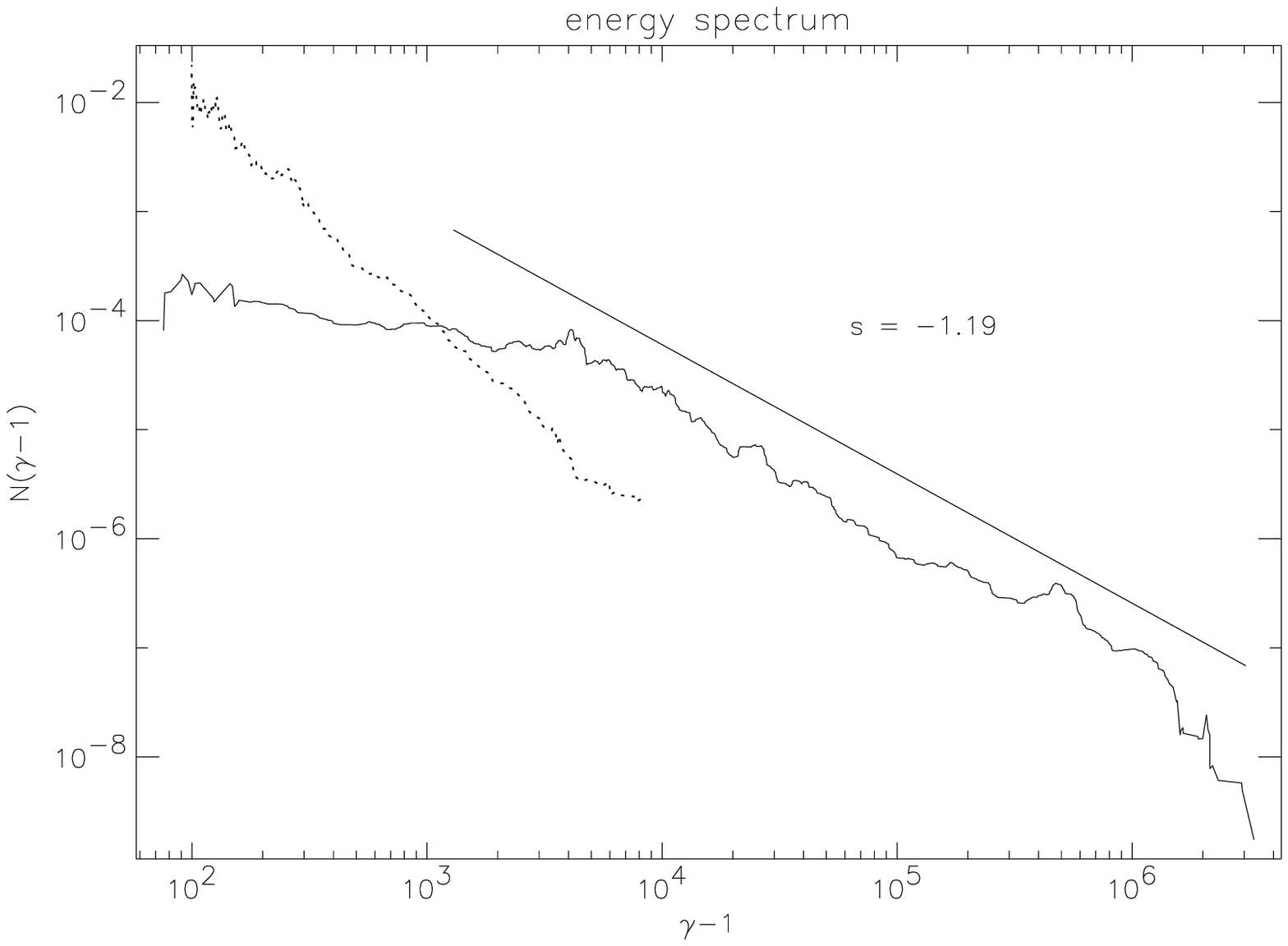}}
    \subfigure[$\overline{\rho} = 0.1 \; m_p/cm^3$]
     {\includegraphics[width=8.5cm,keepaspectratio]{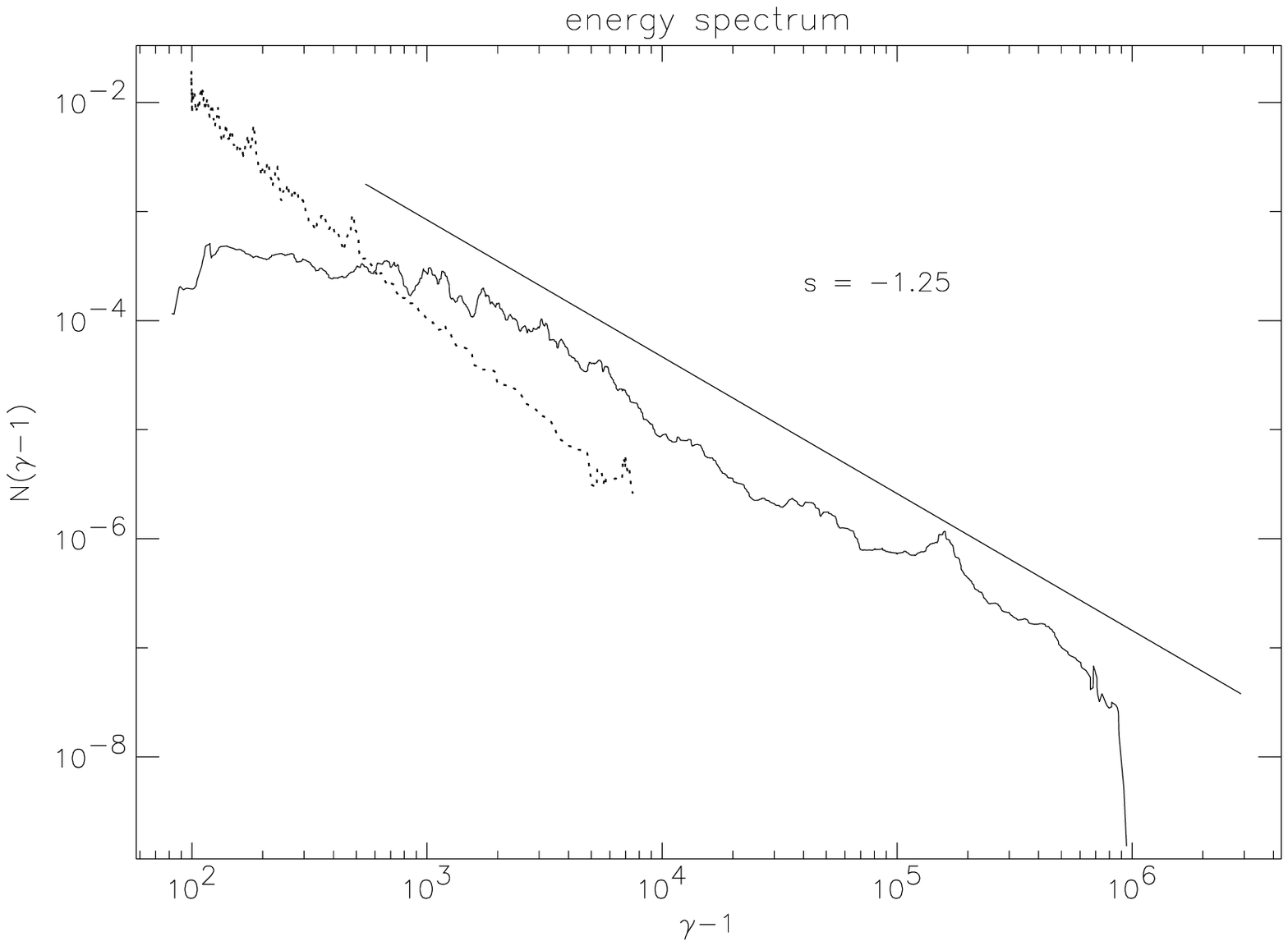}}
  }
  \resizebox{\hsize}{!}{
    \subfigure[$\overline{\rho} = 1 \; m_p/cm^3$]
     {\includegraphics[width=8.5cm,keepaspectratio]{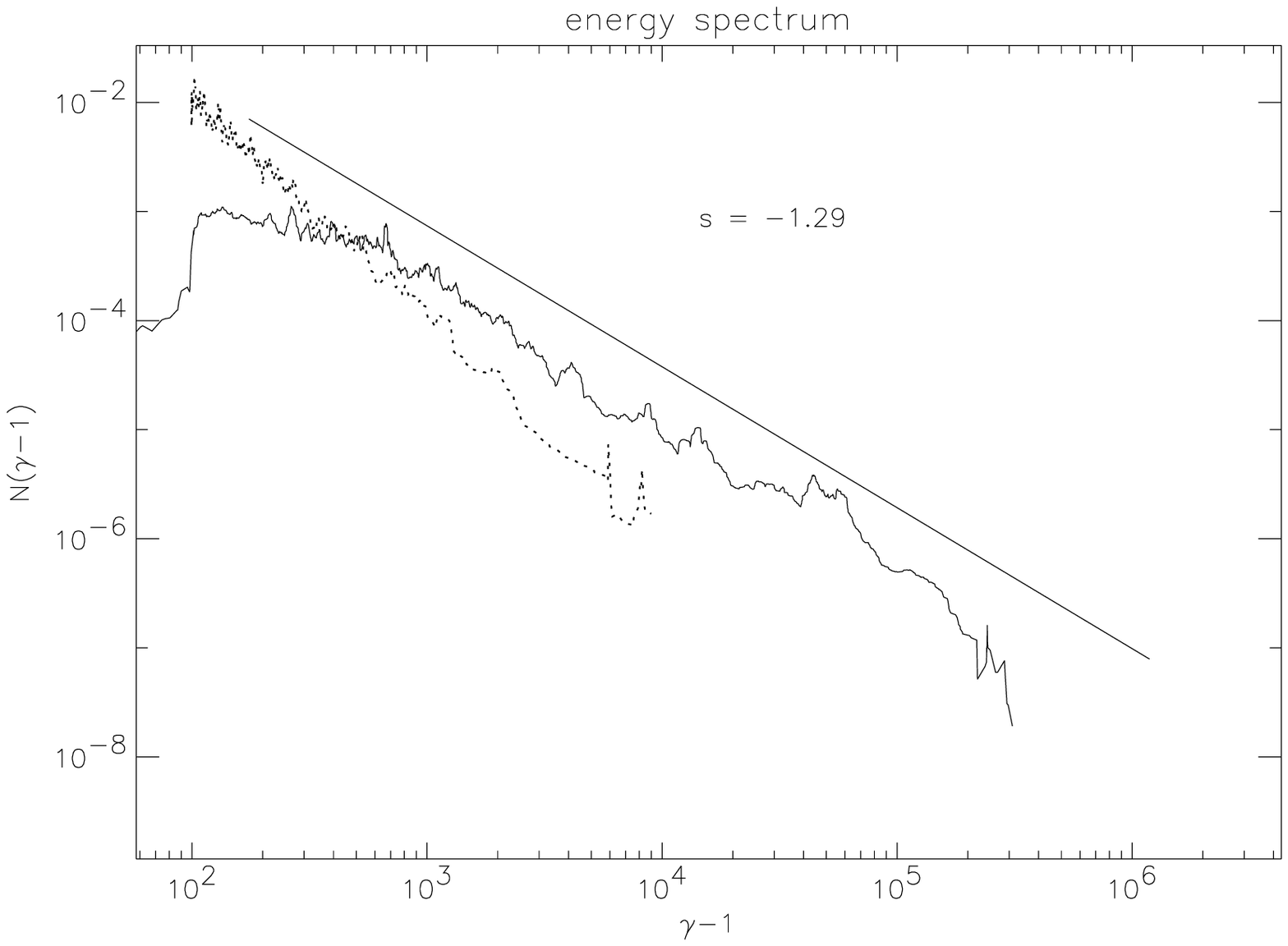}}
    \subfigure[$\overline{\rho} = 10 \; m_p/cm^3$]
     {\includegraphics[width=8.5cm,keepaspectratio]{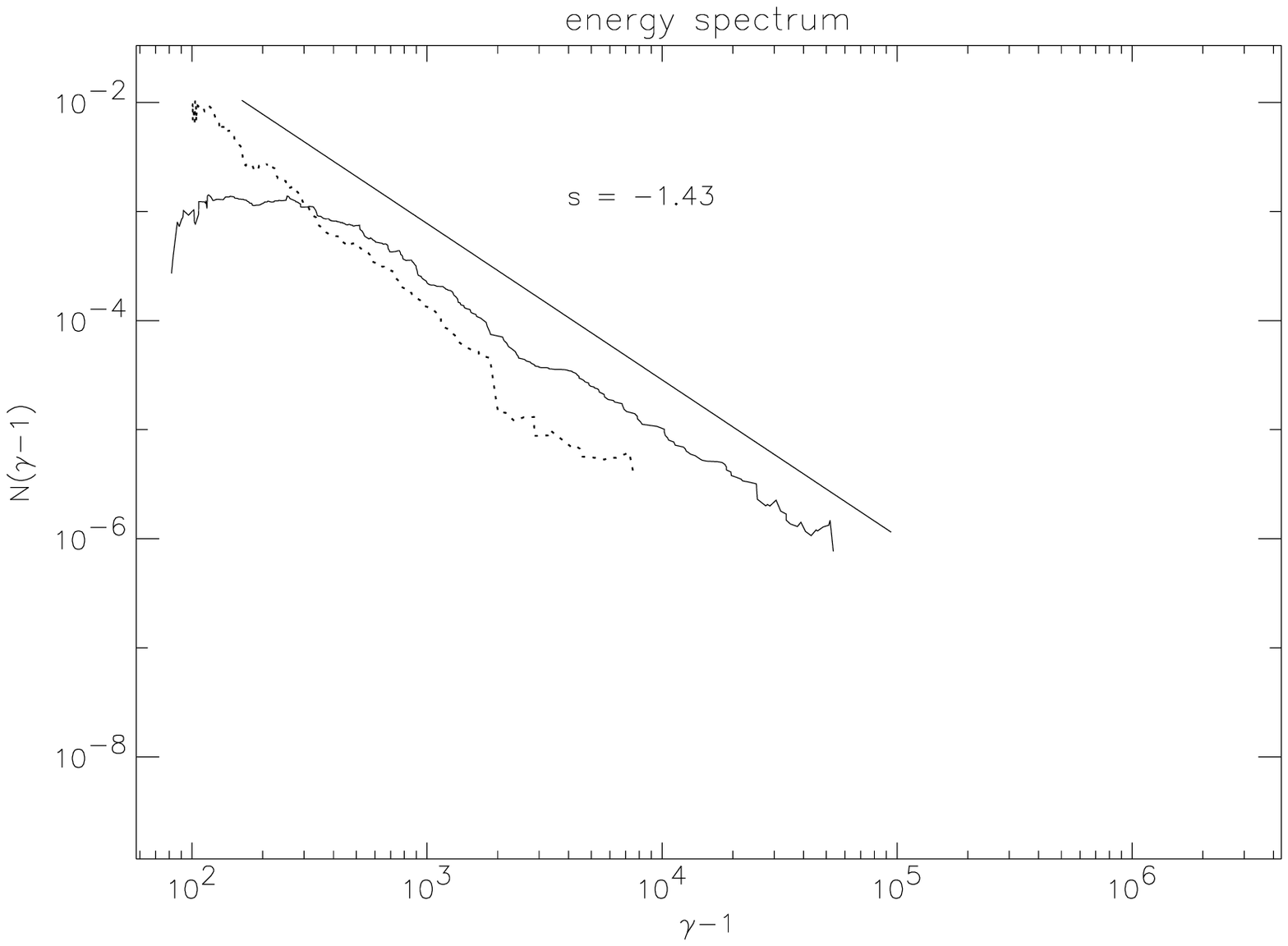}}
  }
  \resizebox{\hsize}{!}{
    \subfigure[$\overline{\rho} = 100 \; m_p/cm^3$]
     {\includegraphics[width=8.5cm,keepaspectratio]{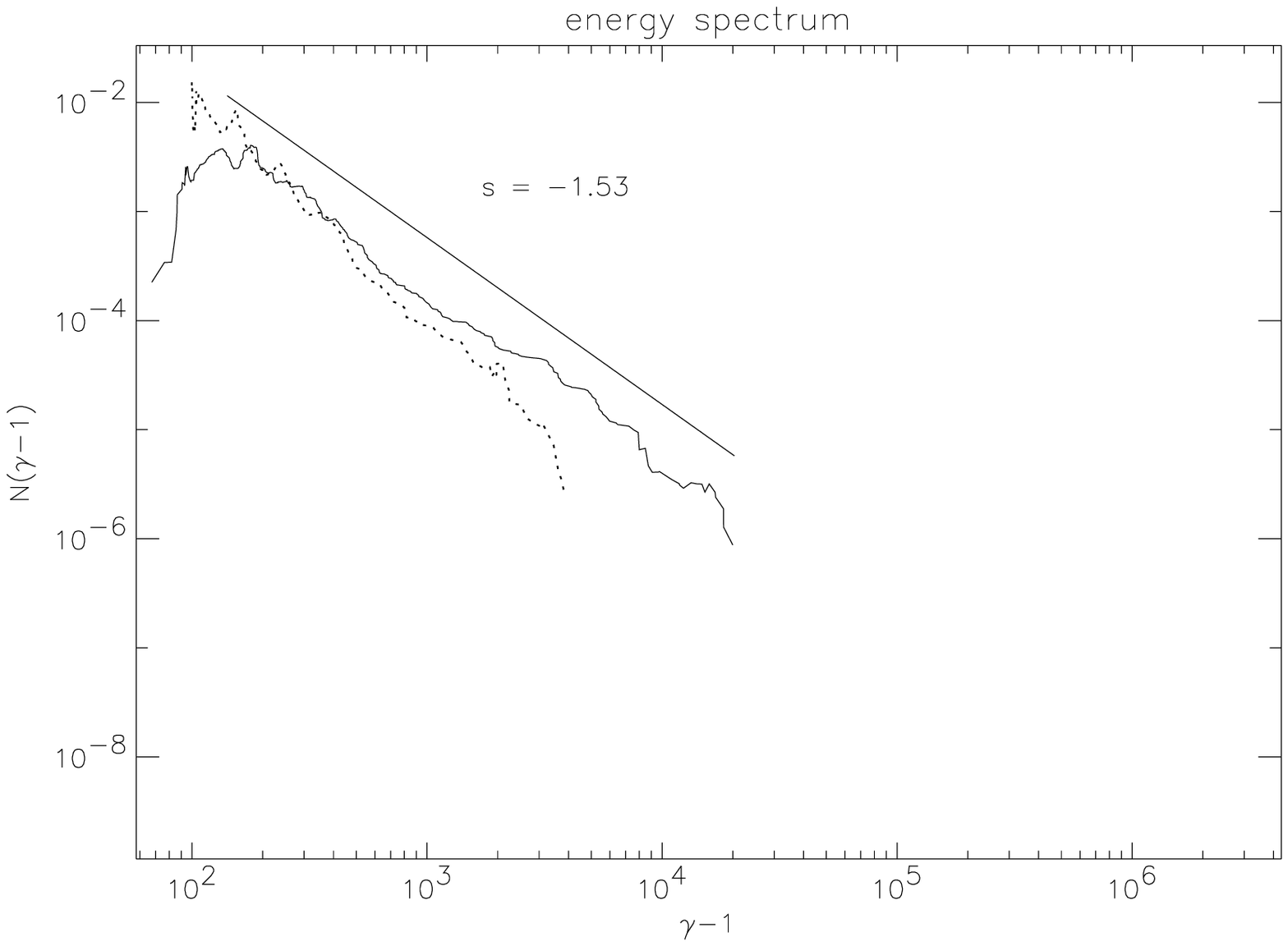}}
    \subfigure[$\overline{\rho} = 1000 \; m_p/cm^3$]
     {\includegraphics[width=8.5cm,keepaspectratio]{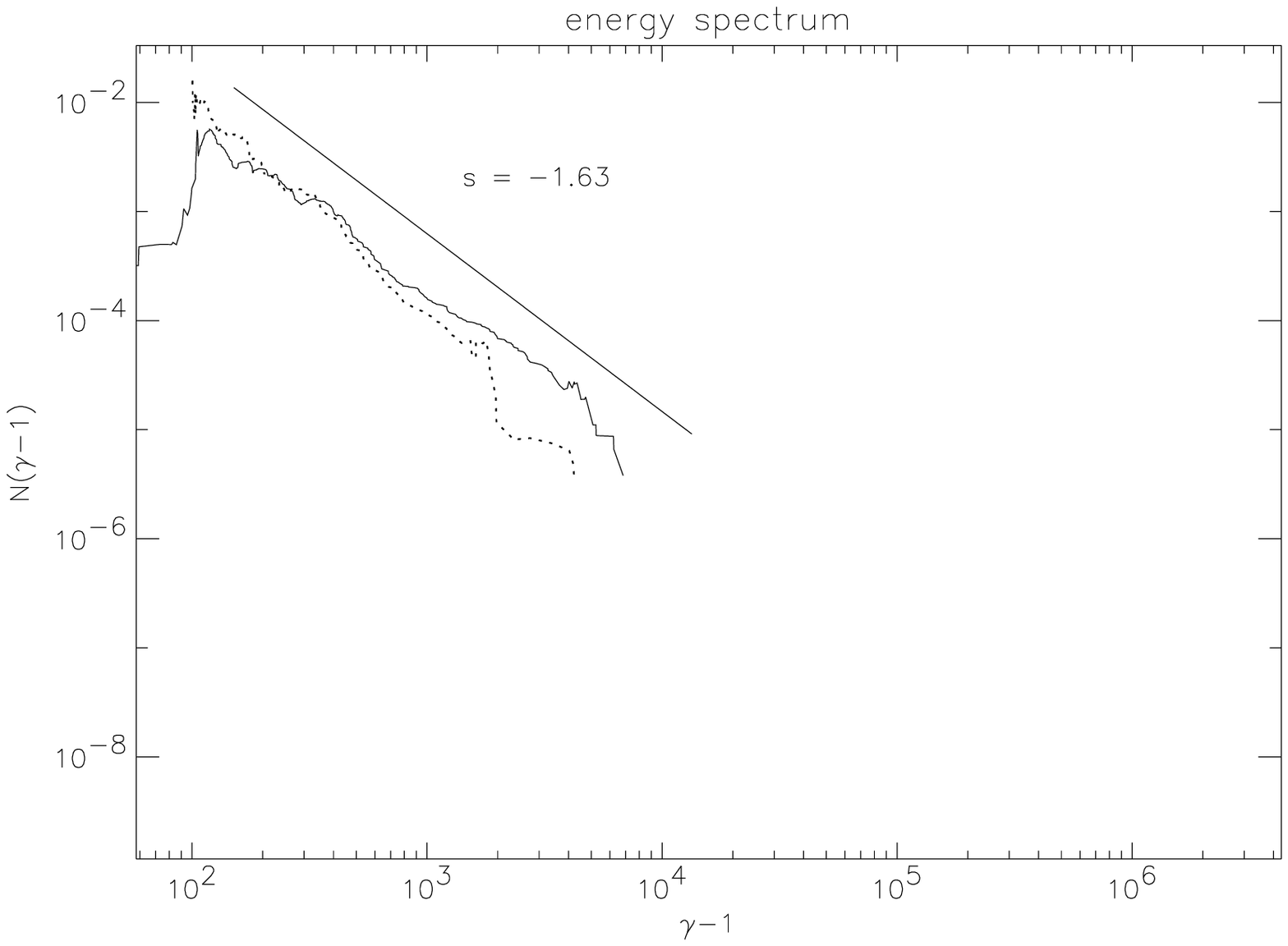}}
  }
  \caption{The initial (dotted line) and final (solid line) energy spectrum of
  the test particle simulations for different values of the mass density
  ($\overline{\rho} = 10^{n} m_p/cm^3$, $n=-2,-1,0,1,2,3$). A least squares fit
  of the spectrum in a certain energy range is plotted above the spectrum and is
  labeled with the spectral index $s$.}
  \label{ortang_energy}
\end{figure*}

Fig. \ref{ortang_energy} shows the initial and the final energy
spectra of the test particle runs for different values of the mean mass
density $\overline{\rho}$. The calculations show that for all values of
$\overline{\rho}$ the spectrum gets flatter than the initial spectrum which
means that particles are accelerated. Compared to the initial spectrum the
low energetic electrons get redistributed to higher energies, i.e. the low
energy peak of the initial spectrum is reduced.

The cut-off energy in the spectrum is correlated with the maximum energy
that the electrons can gain from the mean parallel component of the electric
field $E_{kin}^{max} = e l \overline{E}_{\|}$, where $l$ is the length of
the computational box.

The final energy spectra can be approximated by a broken power law consisting
of two parts: a low energy part which has a relatively flat gradient and
a steeper high energy part. Basically the latter part of the spectrum is
responsible for the synchrotron radiation therefore we will only consider
this part.

For the first run, the one with the lowest density $\overline{\rho}$ we get a
spectral index of $-1.19$ for the high energy part above energies of $\gamma
\approx 4.0\;10^3$. The spectral index $s$ decreases with increasing
$\overline{\rho}$ (approximately by a logarithmic relation), i.e. the
spectrum gets steeper for denser regions. This result corresponds well with
the observed steepening of the radio spectrum at places of magnetic filaments.

\begin{figure*}
  \centering
  \resizebox{\hsize}{!}{
    \subfigure[$\overline{\rho} = 0.01 \; m_p/cm^3$]
     {\includegraphics[width=8.5cm,keepaspectratio]{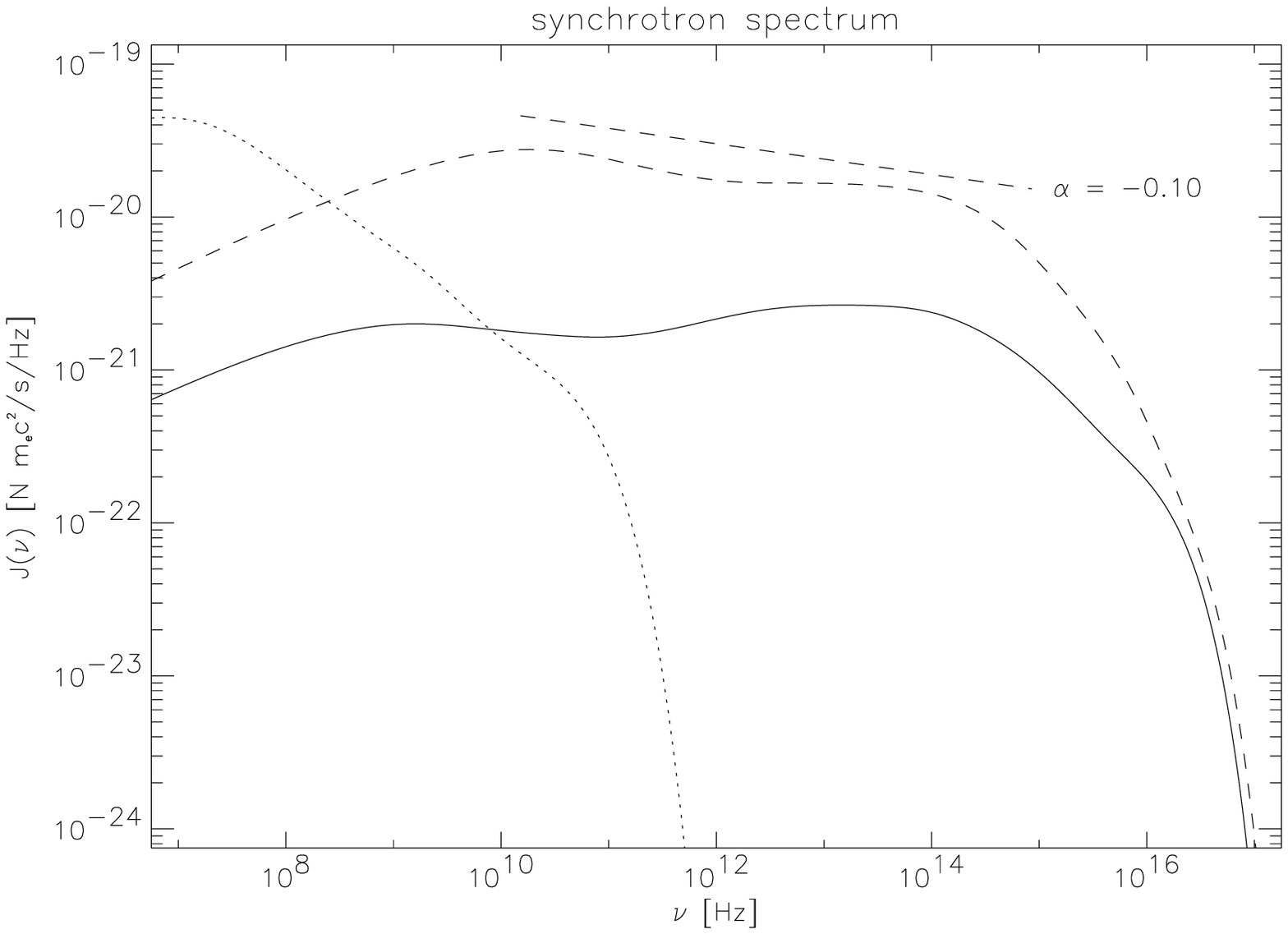}}
    \subfigure[$\overline{\rho} = 0.1 \; m_p/cm^3$]
     {\includegraphics[width=8.5cm,keepaspectratio]{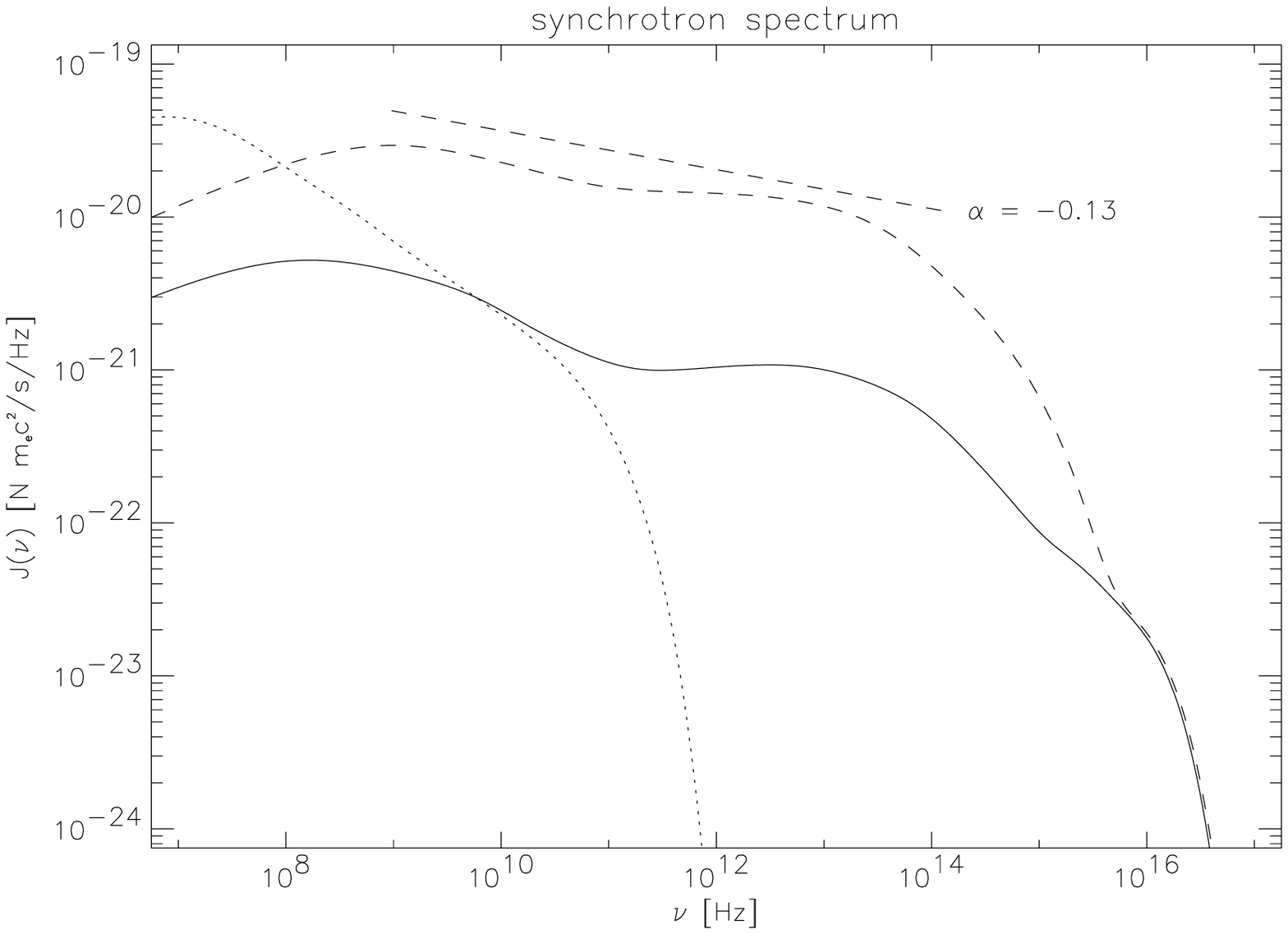}}
  }
  \resizebox{\hsize}{!}{
    \subfigure[$\overline{\rho} = 1 \; m_p/cm^3$]
     {\includegraphics[width=8.5cm,keepaspectratio]{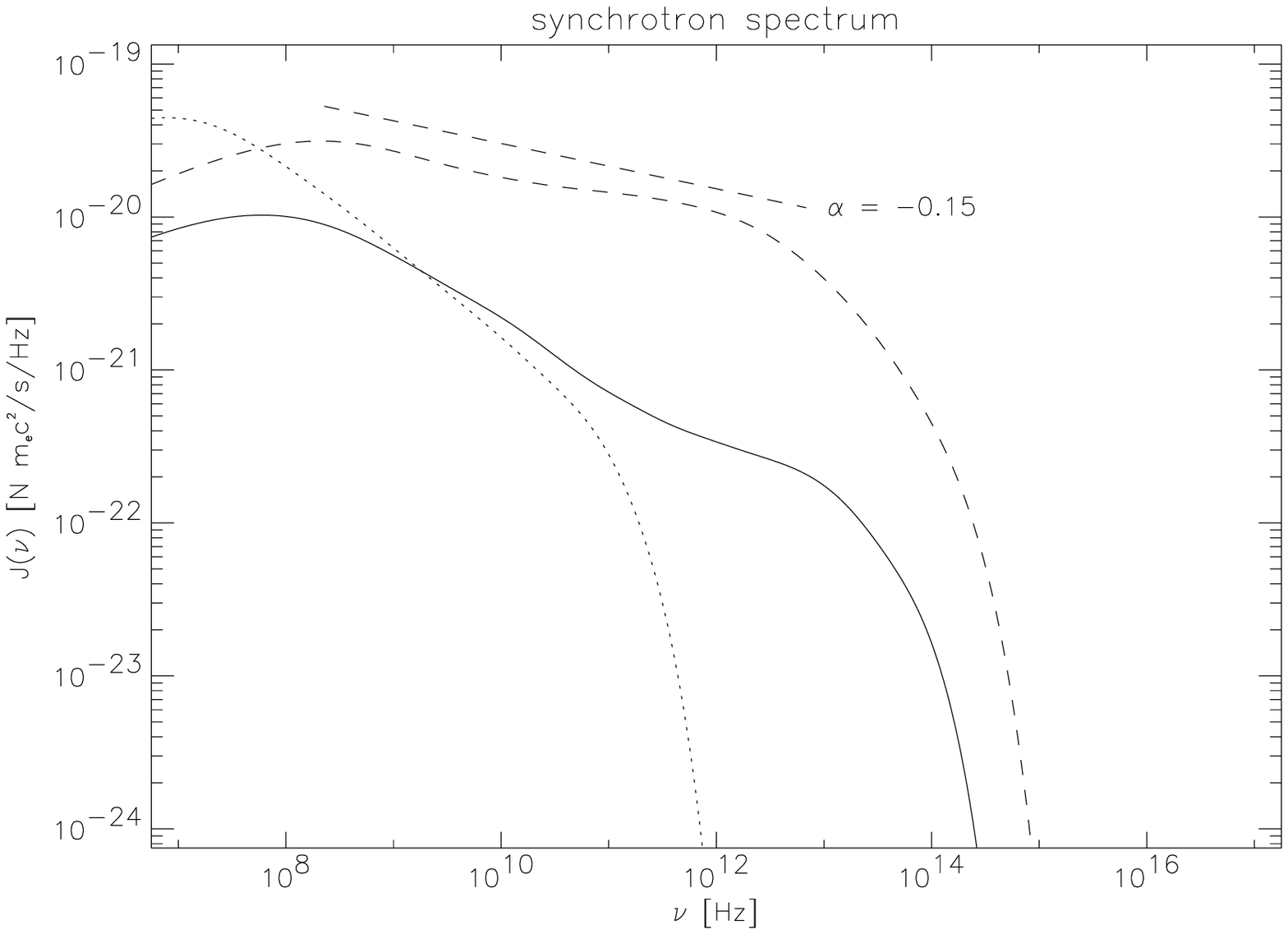}}
    \subfigure[$\overline{\rho} = 10 \; m_p/cm^3$]
     {\includegraphics[width=8.5cm,keepaspectratio]{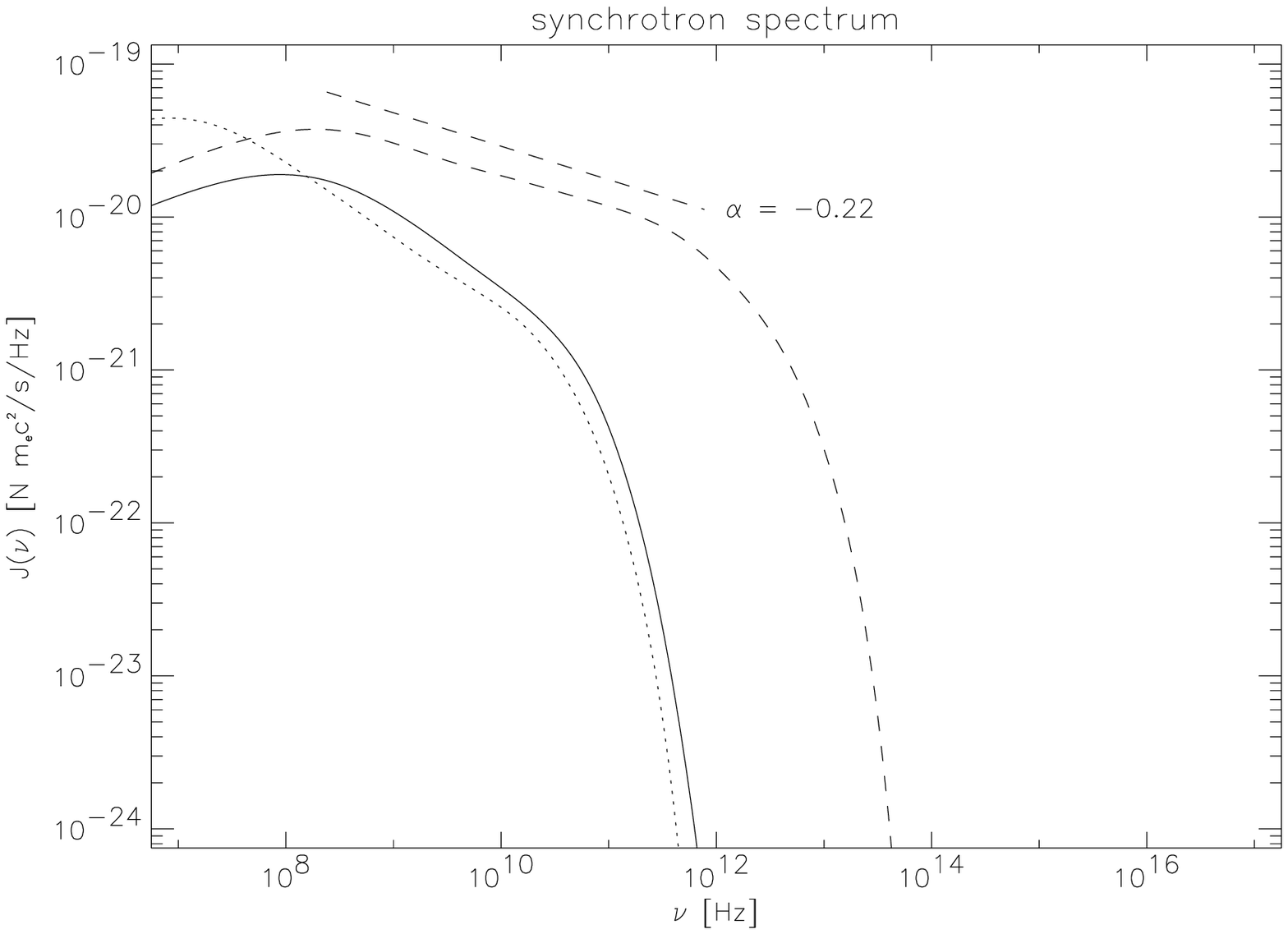}}
  }
  \resizebox{\hsize}{!}{
    \subfigure[$\overline{\rho} = 100 \; m_p/cm^3$]
     {\includegraphics[width=8.5cm,keepaspectratio]{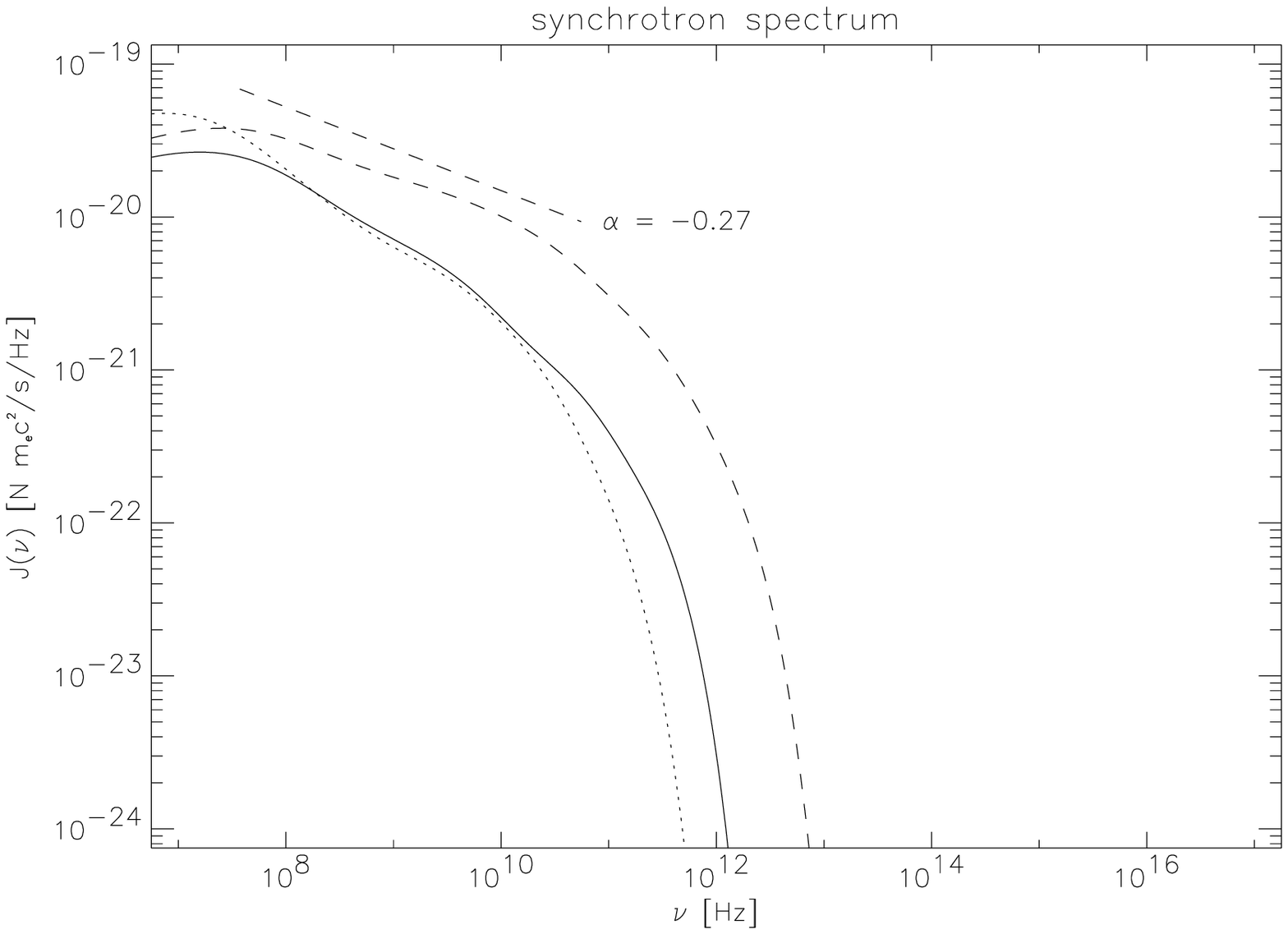}}
    \subfigure[$\overline{\rho} = 1000 \; m_p/cm^3$]
     {\includegraphics[width=8.5cm,keepaspectratio]{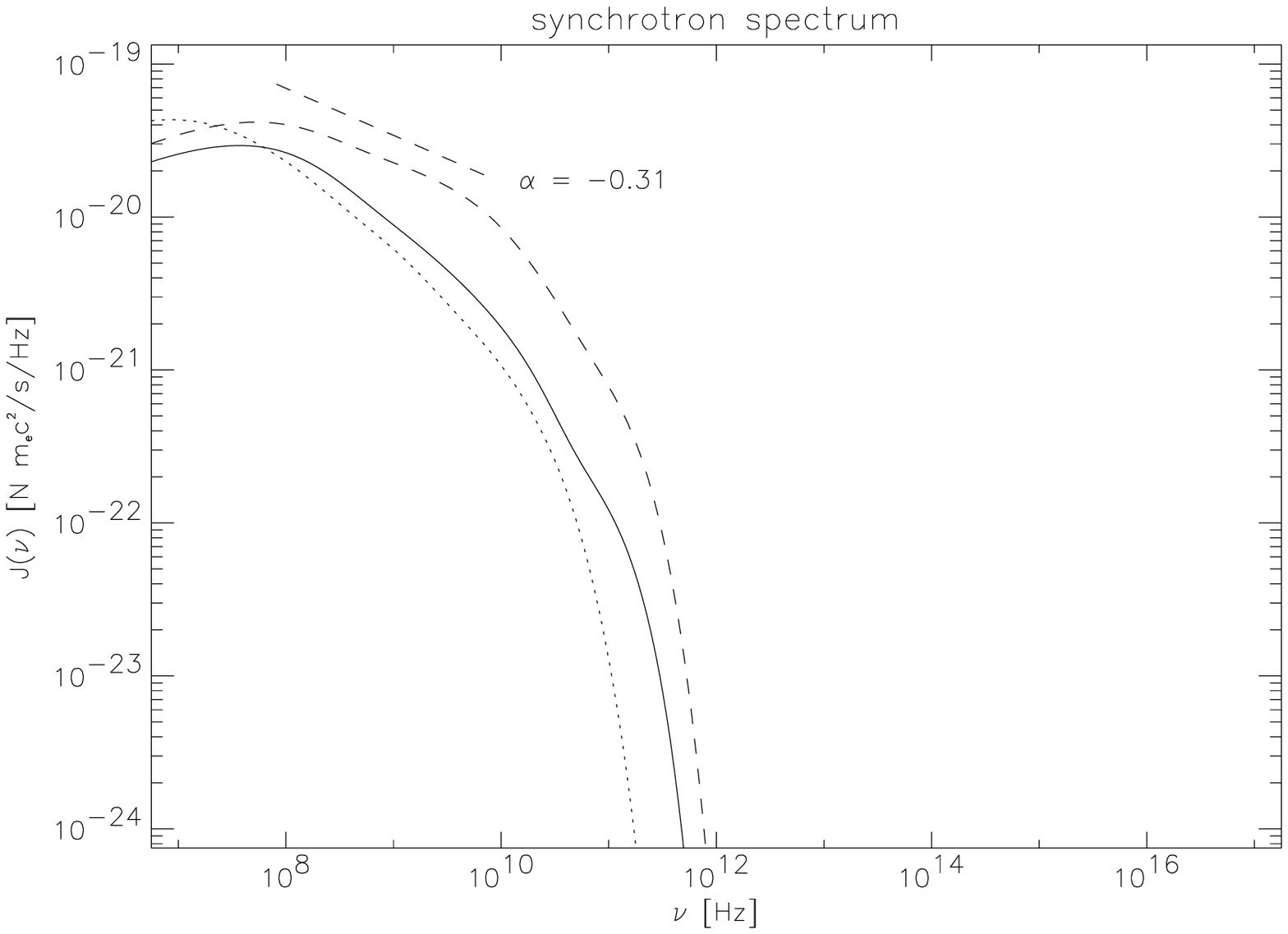}}
  }
  \caption{The initial (dotted line) and the final (solid line) synchrotron
  spectrum of all runs calculated as the sum of the individual synchrotron
  spectra of the corresponding electrons (equation (\ref{power})). The dashed
  line depicts the final synchrotron spectrum of a electron population with
  isotropic pitch angle distribution. The power per unit frequency $J(\nu)$ is
  normalized to the total rest mass energy $N m c^2$ of the set of electrons,
  where $N$ is the number of all electrons.}
  \label{ortang_synch}
\end{figure*}

The resulting synchrotron spectrum can be calculated as the sum of total power
per frequency emitted by each individual electron. The latter is given by
Rybicki \& Lightman \cite{rybicki}
\begin{equation}
  J\left(\frac{\nu}{\nu_c}\right) = 
   \frac{\sqrt{3}e^3}{m c^2} B \sin \theta \frac{\nu}{\nu_c} 
   \int_{\nu/\nu_c}^\infty K_{5/3}(\xi) d\xi,
   \label{power}
\end{equation}
where $\nu_c = \frac{3}{4\pi}\frac{e}{m c} B \gamma^2 \sin\theta$ is the
critical frequency, $\theta$ is the pitch angle and $K_\frac{5}{3}$ is the
modified Bessel function of order $5/3$. Fig. \ref{ortang_synch} shows
the synchrotron spectra of the high energy particles calculated for the
"real" pitch angle distribution taken from our simulation and a isotropic
pitch angle distribution. The isotropic distribution leads to a synchrotron
spectrum with spectral index $\alpha = (s-1)/2$ where $s$ is the index of
the energy spectrum.  This means that the steepening of the energy spectrum
with increasing density results in a steepening of the synchrotron spectrum.

The actual final pitch angle distributions in our simulations are clearly
anisotropic. In the low density runs the pitch angle distributions offer
a peak at $\sin \theta \approx 0.01$. Since $J(\nu) \sim \sin \theta$ the
total radiation power emitted is lower as compared to the isotropic case
(fig. \ref{ortang_synch}). The pitch angle is also correlated with the
gained energy, i.e. the highest energetic particles have the lowest pitch
angles. This correlation results from the acceleration along the magnetic
field lines for which the parallel component of the electric field $E_\| =
\left(\eta \mb{\nabla} \times \mb{B}\right)_{\|}$ is responsible.  $E_\|$ can
only be found at places with significant value of $\eta$. Contrary
to second-order acceleration processes the acceleration due to magnetic
reconnection is able to explain the observed spectra by particle acceleration
in turbulence.  To prove the efficiency of acceleration in presence of
parallel eletric fields we performed a test particle run without any $E_\|$.

\begin{figure*}
  \centering
  \resizebox{\hsize}{!}{
  \includegraphics[width=8.5cm]{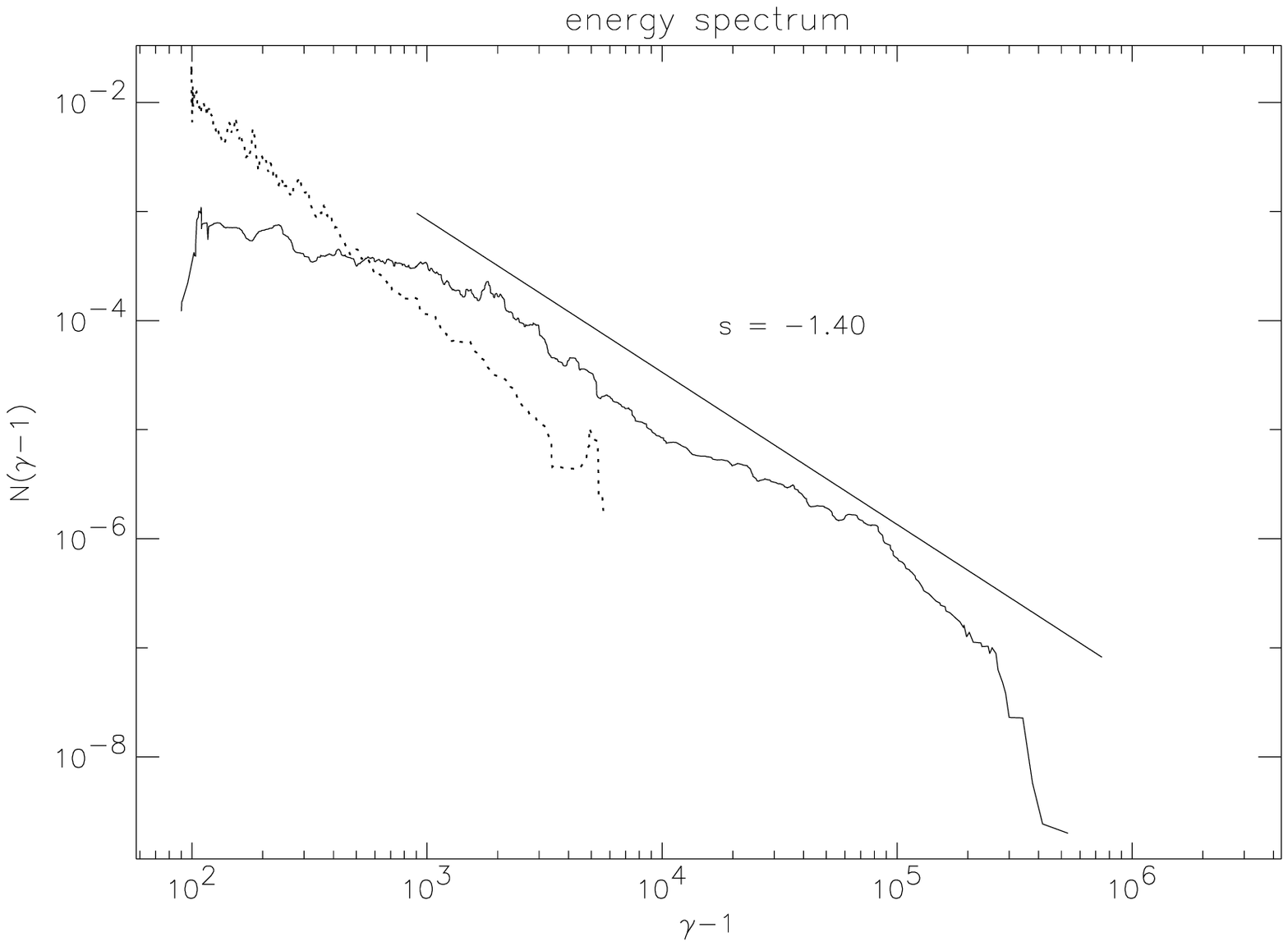}
  \includegraphics[width=8.5cm]{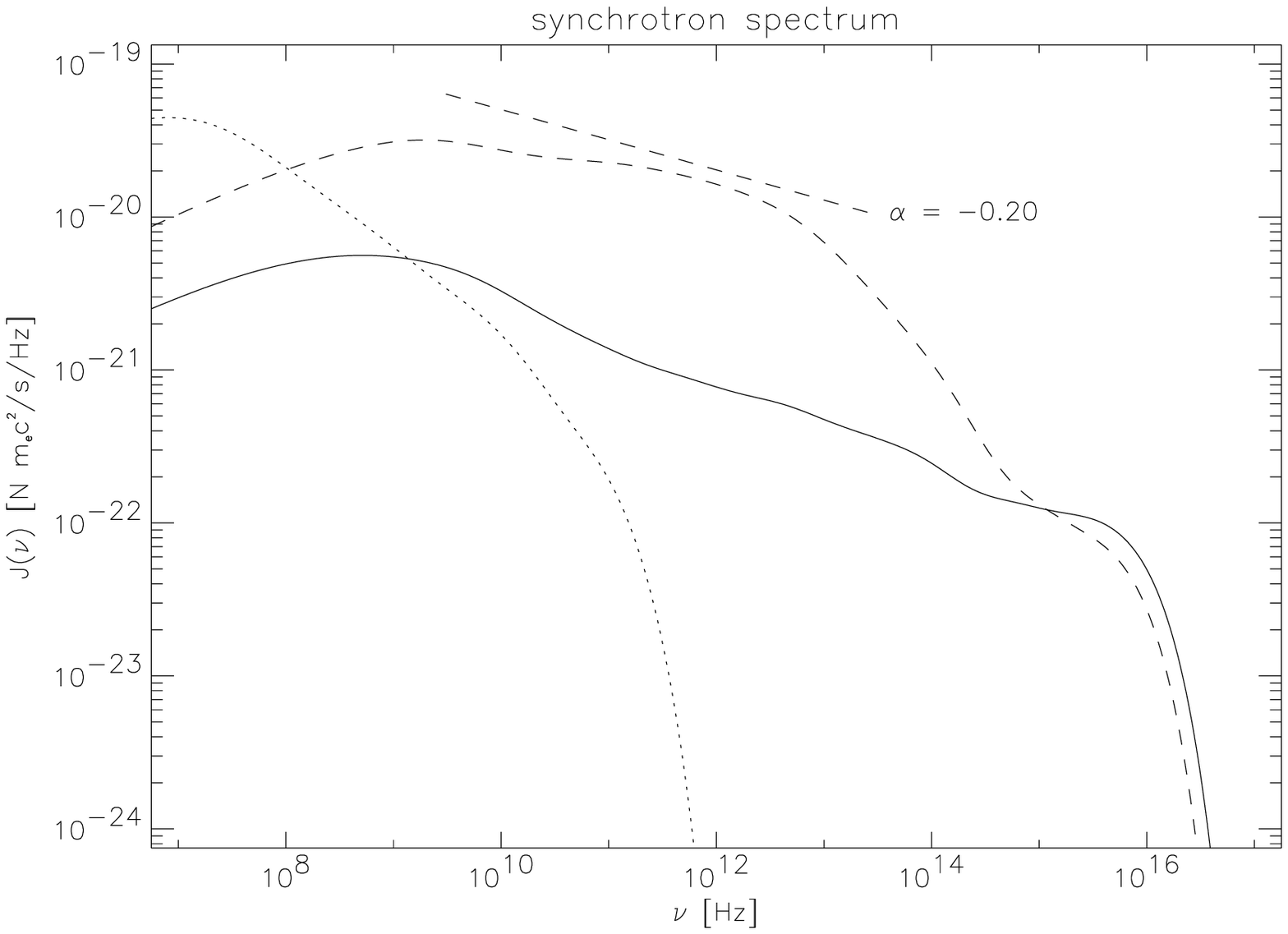}
  }
  \caption{Particle energy and synchrotron spectra of a single run with $E_\| =
  0$. The value for the mean density in this run is the same as in the first run
  in Fig. \ref{ortang_energy} and Fig. \ref{ortang_synch} ($\overline{\rho} =
  10^{-2} m_p/cm^3$). Due to the lower mean electric field energy the 
  achieved energy levels are lower.}
  \label{ortang_2}
\end{figure*}

Fig. \ref{ortang_2} shows the results of a run that is similar to
the first run in Fig. \ref{ortang_energy} and Fig. \ref{ortang_synch}
with the modification $E_\|=0$. The resulting spectra are considerably
steeper, since the acceleration is less efficient as in the case with
reconnection. Therefore the presence of parallel electric fields
is crucial to explain the observed flat spectra. This result is also
corroborated by particle simulations starting from an analytic Kraichnan
type ideal MHD turbulence (these results will be published elsewhere).

\section{Conclusions}
\label{discussion}

\begin{figure}
  \resizebox{\hsize}{!}{
    \includegraphics{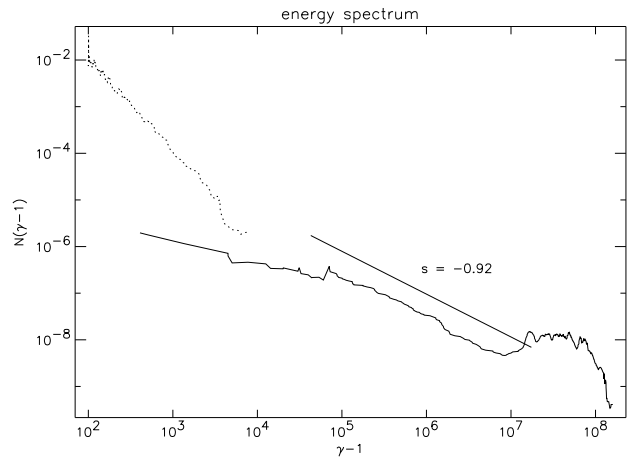}
  }
  \caption{The initial (dotted line) and final (solid line) energy spectrum
  of a test particle run with larger box size of $l = 10^{17} \mr{cm}$. The
  density for this run was chosen to be $\hat{\rho} = 10^{-2} m_p/cm^3$.}
  \label{ortang01c}
\end{figure}

We studied the acceleration of relativistic electron populations in
three-dimensional reconnective turbulence with applications to the Crab
nebula. By means of test particle simulations performed within
non-linearly evolved electromagnetic fields modeled by resistive MHD
simulations and including radiative losses via synchrotron emission we could
show that particles are accelerated and form a energy distribution with
powerlaw index $\alpha$ ranging from $1.2$ to $1.6$. We can give limits for
the maximum size of the turbulent regions in the nebulae. If the volume is too
large and the particle escape time is too long their spectra get significantly
harder, i.e. the spectral index is flatter than $1$ (see Fig. \ref{ortang01c}).
For the case of the Crab nebula we can rule out turbulent cells larger than
$10^{16}\rm cm$. This limit is in accordance with experimental results from
turbulent fluids, in which the turbulence size is limited by the spatial scale
of the turbulence source. Since the Crab nebula is powered by a pulsar wind
which is shocked at about $10^{17}\rm cm$ the spatial scale of the turbulence
should be smaller.

We like to suppose the following scenario for the radio emission of the Crab
nebula: The nebula is permanently powered by the plasma wind originating
from the pulsar. Consequently strong MHD Turbulence is exited in the
nebula. In the turbulent regions the leptons experience in situ acceleration
in the reconnective turbulent environment. Our simulations prove that the
resulting energy spectra and synchrotron spectra are considerably flat and
therefore can explain the observed hard radio spectra (Weiler \& Shaver
\cite{weiler2}). Whereas we concentrate on the Crab nebula, our findings
should also be applicable to flat radio spectra of plerions in general.
We note, that the efficency of the acceleration within th test particle
approach depends on the ratio of Alfven speed to velocity of light which
determines the effective electric field strength responsible for particle
energization. If the Alfven speed decreases since the magnetic field decreases
the electric field will also be weaker, reducing the acceleration to higher
energies.  If a supernova remnant consists of region with significantly varying
magnetic fields and densities the nonthermal particles may be a superposition
of particles experiencing different efficencies and electromagnetic fields.

We note that Atoyan (\cite{atoyan}) considered the origin of radio emitting
electrons in the Crab nebula in terms of intensive radiative and adiabatic
cooling in the past. Whereas he argued that effective in situ acceleration in
the nebula is not necessary to explain the radio spectra, our simulations prove
the capability of magnetic reconnection to energize an electron population
of a magnetized plasma under the influence of external electromagnetic fields
onto which the accelerated leptons have no back reaction.

\end{document}